\documentclass[10pt,twoside]{article}
\usepackage{amssymb}

\usepackage{graphicx}

\begin{document}

\title{Bayesian Analysis of the Chaplygin Gas and Cosmological Constant Models
using the SNe Ia Data}
\author{R. Colistete Jr.\thanks{%
e-mail: \texttt{colistete@cce.ufes.br}}, J. C. Fabris\thanks{%
e-mail: \texttt{fabris@cce.ufes.br}}, S.V.B. Gon\c{c}alves\thanks{%
e-mail: \texttt{sergio@cce.ufes.br}} \, and P.E. de Souza\thanks{%
e-mail: \texttt{patricia.ilus@bol.com.br}} \\
\\
\mbox{\small Universidade Federal do Esp\'{\i}rito Santo,
Departamento
de F\'{\i}sica}\\
\mbox{\small Av. Fernando Ferrari s/n - Campus de Goiabeiras, CEP
29060-900, Vit\'oria, Esp\'{\i}rito Santo, Brazil}}
\date{\today}
\maketitle

\begin{abstract}
The type Ia supernovae observational data are used to estimate the
parameters of a cosmological model with cold dark matter and the Chaplygin
gas. This exotic gas, which is characterized by a negative pressure varying
with the inverse of density, represents in this model the dark energy
responsible for the acceleration of the Universe. The Chaplygin gas model
depends essentially on four parameters: the Hubble constant, the velocity of
the sound of the Chaplygin gas, the curvature of the Universe and the
fraction density of the Chaplygin gas and the cold dark matter. The Bayesian
parameter estimation yields $H_0 = 62.1^{+3.3}_{-3.4}\,km/M\!pc.s$, $%
\Omega_{k0} = -0.84^{+1.51}_{-1.23}$, $\Omega_{m0} = 0.0^{+0.82}_{-0.0}$, $%
\Omega_{c0} = 1.40^{1.15}_{-1.16}$, $\bar{A} = c_s^2 =
0.93^{+0.07}_{-0.21}\,c $, $t_0 = 14.2^{+2.8}_{-1.3}\,Gy$ and $q_0
= - 0.98^{+1.02}_{-0.62}$. These and other results indicate that a
Universe completely dominated by the Chaplygin gas is favoured,
what reinforces the idea that the Chaplygin gas may unify the
description for dark matter and dark energy, at least as the type
Ia supernovae data are concerned. A closed and accelerating
Universe is also favoured. The Bayesian statistics indicates that
the Chaplygin gas model is more likely than the standard
cosmological constant ($\Lambda CDM$) model at $55.3\%$ confidence
level when an integration on all free parameters is performed.
Assuming the spatially flat curvature, this percentage mounts to
$65.3\%$. On the other hand, if the density of dark matter is
fixed at zero value, the Chaplygin gas model becomes more
preferred than the $\Lambda CDM$ model at $91.8\%$ confidence
level. Finally, the hypothesis of flat Universe and baryonic
matter ($\Omega _{b0}=0.04$) implies a Chaplygin gas model
preferred over the $\Lambda CDM$ at a confidence level of
$99.4\%$. \vspace{0.7cm}
\end{abstract}

PACS number(s): 98.80.Bp, 98.65.Dx \vspace{0.7cm}

\section{Introduction}

The combined data from the measurements of the spectrum of anisotropies of
the cosmic microwave background radiation \cite{charles} and from the
observations of high redshift type Ia supernovae \cite{riess,perlmutter}
indicate that the matter content of the Universe today may be very probably
described by cold dark matter and dark energy, in a proportion such that $%
\Omega _{dm}\approx 0.3$ and $\Omega _{de}\approx 0.7$. These dark
components of the matter content of the Universe manifest themselves only
through their gravitational effects. At same time, a fraction of the dark
components agglomerates at small scales (cold dark matter) while the other
fraction seems to be a smooth component (dark energy). The dark energy must
exhibit negative pressure, since it would be the responsible for the present
acceleration of the Universe as indicated by the type Ia supernovae
observations, while the cold matter must have zero (or almost zero)
pressure, in order that it can gravitationally collapse at small scales.

The nature of these mysterious matter components of the Universe is object
of many debates. The cold dark matter may be, for example, axions which
result from the symmetry breaking process of Grand Unified Theories in the
very early Universe. But, since Grand Unified Theories, and their
supersymmetrical versions, remain a theoretical proposal, the nature of cold
dark matter is yet an open issue.

A cosmological constant is, in principle, the most natural candidate to
describe the dark energy. It contributes with an homogeneous, constant
energy density, its fluctuation being strictly zero. However, if the origin
of the cosmological constant is the vacuum energy, there is a discrepancy of
about $120$ orders of magnitude between its theoretical value and the
observed value of dark energy \cite{carroll}. This situation can be
ameliorate, but not solved, if supersymmetry is taken into account. Another
candidate to represent dark energy is quintessence, which considers a
self-interacting scalar field, which interpolates a radiative era and a
vacuum dominated era \cite{steinhardt,sahni,brax}. But the quintessence
program suffers from fine tuning of microphysical parameters \cite{lyth}.

Recently, an alternative to both the cosmological constant and to
quintessence to describe dark energy has been proposed: the Chaplygin gas
\cite{pasquier,patricia,bilic,bento}. The Chaplygin gas is characterized by
the equation of state
\begin{equation}  \label{eoe}
p = - \frac{A}{\rho} \quad ,
\end{equation}
where $A$ is a constant. Hence, the pressure is negative while the sound
velocity is positive, avoiding instability problems at small scales \cite
{jerome}. The Chaplygin gas has been firstly conceived in studies of
adiabatic fluids \cite{chaplygin}, but recently it has been identified an
interesting connection with string theories \cite{hoppe,ogawa,jackiw}. Some
extensions of the Chaplygin gas have been proposed \cite{bento} through the
equation of state
\begin{equation}  \label{geoe}
p = - \frac{A}{\rho^\alpha} \quad .
\end{equation}
However, the connection with string theories is lost unless $\alpha = 1$,
and for this reason we will only consider in what follows the equation of
state (\ref{eoe}).

Considering a relativistic fluid with the equation of state (\ref{eoe}), the
equation for the energy-momentum conservation relations leads, in the case
of an homogeneous and isotropic Universe, to the following relation between
the fluid density and the scale factor $a$:
\begin{equation}
\rho =\sqrt{A+\frac{B}{a^{6}}}\quad ,
\end{equation}
where $B$ is an integration constant. This relation shows that initially the
Chaplygin gas behaves as a pressureless fluid, acquiring later a behaviour
similar to a cosmological constant. So, it interpolates a decelerated phase
of expansion to an accelerated one, in way close to that of the quintessence
program.

In this work, we will constrain the parameters associated with the Chaplygin
gas model ($CGM$) using the type Ia supernovae (SNe Ia) observational data.
Specifically, we will consider a model where the dynamics of the Universe is
driven in principle by pressureless matter and by the Chaplygin gas. The
luminosity distance for the configuration where, in general, Chaplygin gas
and dark matter are present, is evaluated from which the relation between
the magnitude and the redshift $z$ is established. The observational data
are then considered, and they are fitted using four free parameters: the
ratio of the density fraction, with respect to the critical density $\rho
_{c}$, of the pressureless matter and of the Chaplygin gas, $\Omega _{m0}$
and $\Omega _{c0}$ respectively, the sound velocity of the Chaplygin gas $%
\bar{A}$, in terms of the velocity of light, the curvature parameter $\Omega
_{k0}$ and the Hubble parameter $H_{0}$. All these parameters are evaluated
today.

The fact that we consider a model containing pressureless matter and the
Chaplygin gas, implies that in principle we ignore the possibility that the
Chaplygin gas could unify the description of dark energy and dark matter
\cite{bilic,bento}. This unification is suggested by the fact that, at
perturbative level, the Chaplygin gas does not agglomerate at very large
scales while it may agglomerate at small scales. The possible unification of
both dark components of the matter content of the Universe through the
Chaplygin gas has increased the interest on this new possible exotic fluid.
A recent analysis of type Ia supernovae data correlated with the $CGM$ \cite
{martin} used extensively this idea, employing the generalized equation of
state (\ref{geoe}) in a flat Universe and excluding, \textit{ab initio}, the
possibility to have a dark matter component. In this work, the authors tried
to restrict the possible values of the parameter $\alpha $, but their
results indicate a yet large range of possible values for this parameter. In
the Ref. \cite{avelino}, the authors studied, using type Ia supernovae data,
essentially two scenarios: a flat Universe with the generalized Chaplygin
gas and dark matter; a non-flat Universe with the ordinary Chaplygin gas and
baryonic matter (with a fixed $\Omega _{b}=0.04$). For the last case, the
most favoured configuration was obtained for a closed Universe with $\Omega
_{c0}=1.27$ and $\bar{A}=0.78$, using a $\chi ^{2}$ statistic. As it will be
seen later, this result is consistent with the more general analysis we will
perform here.

The unification program using the Chaplygin gas has been recently criticized
\cite{ioav,bean}. In our opinion this question remains an open issue, for
the following reasons. In Ref. \cite{ioav}, the authors consider a model
containing only the Chaplygin gas, and find a spectrum for the clustering of
matter that is not in agreement with the 2dFGRS observations, unless the
parameter $\alpha $ in (\ref{geoe}) is very small, $\alpha <10^{-5}$.
However, we must not forget that there are baryons in the Universe. Even the
quantity of baryons is small, it presences claims at least for a two fluid
model, whose behaviour is in general quite different of a single fluid model
\cite{glauber}. Moreover, we may rise the question of what is really
observed in the matter power spectrum. In Ref. \cite{bean}, the authors are
more stringent by stating that, when the large scale structure data are
crossed with the CMB data, the likelihood in the configuration space leads
to a range of value for the $\alpha $ parameter that is very near zero and $%
\bar{A}\sim 1$, i.e., essentially a cosmological constant model. However,
the CMB analysis suffers from a high degeneracy in the space of the
parameters, and the crossing of these data with those coming from type Ia
supernovae must bring again the Chaplygin gas to scene. Here, we open the
possibility to have dark matter present, not excluding at the same time the
possibility to have a pure $CGM$.

Our goal here, in contrast with the Refs. \cite{martin,avelino}, is to
perform an extensive and comprehensive analysis of the problem of fitting
the type Ia supernovae data using the ordinary Chaplygin gas given by (\ref
{eoe}). This restriction on the parameter $\alpha $ keeps us in contact with
a string motivation for the Chaplygin gas. At the same time, our intention
is to leave space for the presence of other fluids besides the Chaplygin gas
as well as a spatial curvature term, constraining these parameters (together
with the Chaplygin gas sound velocity and the Hubble parameter) only through
the confrontation with the supernovae type Ia data. In this sense, the
present work is more general than the previous ones, excepting for the
restriction on the equation of state, restriction dictated by the
theoretical motivation for the Chaplygin gas. A detailed description of the
Bayesian statistics method of analysis is presented, as well as the meaning
of the observational limits on the different free parameters of the model
obtained through this method. A comparison with the cosmological constant
model ($\Lambda CDM$) is exhibited, which is one of the main purpose of this
work.

It will be shown, in particular, that the $CGM$, in what concerns the
supernovae type Ia data, favours strongly, at $2\sigma $, a closed Universe (%
$\Omega _{k0}=-0.84_{-1.23}^{+1.51}$), peaked in the zero value for the cold
dark matter density ($\Omega _{m0}=0.0_{-0.0}^{+0.82}$) and a sound velocity
$c_{s}^{2}=0.93_{-0.21}^{+0.07}\,c$ near, but not equal, to the value that
corresponds to the cosmological constant. The present age of the Universe is
$t_{0}=14.2_{-1.3}^{+2.8}\,Gy$ in the $CGM$, a bit smaller than that in the $%
\Lambda CDM$ model ($t_{0}=15.4_{-1.9}^{+3.4}\,Gy$), but still compatible
with other astronomical measurements \cite{krauss}. The deceleration
parameter is highly negative: $q_{0}=-0.98_{-0.62}^{+1.02}$. The $CGM$ is
always prefered with respect to the $\Lambda CDM$ model, with a confidence
level of $55.3\%$ when all free parameters are considered. This confidence
level is considerably higher if the curvature or the dark matter density is
fixed to zero. An almost flat Universe is predicted if dark mater is absent,
$\Omega _{k0}=0.17_{-1.58}^{+0.83}$($\Omega _{m0}=0$). Later, we will make a
brief comparison between the results obtained with the recent data coming
from the anisotropy of the cosmic microwave background (including the data
coming from the WMAP) \cite{spergel,melchiorri}, 2dFGRS measurements \cite
{colless} and Lyman $\alpha $ forest \cite{forest}, even if a proper
combined analysis is outside the scope of the present paper.

In next section, the Chaplygin gas model is described. In section 3, the
Bayesian probabilistic analysis is described. In section 4, the cosmological
parameters are estimated, and in section 5 our conclusions are presented.

\section{The Chaplygin gas model and the least-squares fitting}

We now proceed by constructing the cosmological model based on the Chaplygin
gas, which we will test against the type Ia supernovae observations. We must
specify the observable cosmological parameters in the model: we will deal
with the Hubble parameter $H_{0}$, the curvature parameter $\Omega _{k0}$,
the dark matter density parameter $\Omega _{m0}$, the Chaplygin gas density
parameter $\Omega _{c0}$ and the Chaplygin gas sound velocity $c_{s}^{2}$.
All these parameters are evaluate today.

The sound velocity of the Chaplygin gas today, in units of $c$, is given by
\begin{equation}
c_{s}^{2}=\bar{A}=\frac{A}{\rho _{c0}^{2}}\quad .
\end{equation}
The ratios of the density fractions with respect to the current critical
density are related by $\Omega _{k0}+\Omega _{m0}+\Omega _{c0}=1$. For the
Chaplygin gas case we take $\Omega _{m0}\geqslant 0$ and $\Omega
_{c0}\geqslant 0$, so $\Omega _{k0}\leqslant 1$. When the limit $\bar{A}%
\rightarrow 1$ is taken ($c=1$ hereafter), the cosmological constant model
is obtained from the Chaplygin model, so $\Omega _{c0}$ is relabeled as $%
\Omega _{\Lambda }$ (the ratio of the vacuum energy density with respect to
the current critical density), and the only range restriction is $\Omega
_{m0}\geqslant 0$. It will be verified that the best-fitting and the
Bayesian inference suggest $\Omega _{m0}\approx 0$, this result becomes
quite interesting if we take into account some recent considerations about a
unification of cold dark matter and dark energy in Chaplygin gas models \cite
{bento}. Moreover, closed models will be favoured.

The equation governing the evolution of our model is
\begin{equation}
\left( \frac{\dot{a}}{a}\right) ^{2}+\frac{k}{a^{2}}=\frac{8\pi G}{3}\left(
\frac{\rho _{m0}}{a^{3}}+\sqrt{A+\frac{B}{a^{6}}}\right) \quad \quad .
\end{equation}
It can be rewritten as
\begin{equation}
\left( \frac{\dot{a}}{a}\right) ^{2}=H_{0}^{2}\left( \frac{\Omega _{m0}}{%
a^{3}}+\Omega _{c0}\sqrt{\bar{A}+\frac{1-\bar{A}}{a^{6}}}+\frac{\Omega _{k0}%
}{a^{2}}\right)  \label{H2b}
\end{equation}
where $H_{0}$ is the Hubble parameter today, $\Omega _{i}=\frac{\rho _{i}}{%
\rho _{c}}$ ($\rho _{i}$ denoting a matter component and $\rho _{c}$ the
critical density), $\Omega _{k0}=\frac{3kc^{2}}{8\pi G}$ and the scale
factor was normalized to unity today, $a_{0}=1$.

The luminosity distance is obtained by standard procedures \cite{coles},
using the equation for the light trajectory in the above specified
background, and its definition,
\begin{equation}
D_{L}=\left( \frac{1}{4\pi }\frac{L}{l}\right) ^{1/2}
\end{equation}
where $L$ is the absolute luminosity of the source, and $l$ is the
luminosity measured by the observer. This expression can be rewritten as
\begin{equation}
D_{L}=(1+z)r\quad ,
\end{equation}
$r$ being the co-moving distance of the source. Taking into account the
definitions of absolute and apparent magnitudes in terms of the luminosity $%
L $ and $l$, $M$ and $m$ respectively, we finally obtain the relation valid
for the three types of spatial section:
\begin{eqnarray}
m &-&M=5\log \left\{ (1+z)\frac{c}{H_{0}}\frac{1}{\sqrt{\left| \Omega
_{k0}\right| }}\times f\left[ \sqrt{\left| \Omega _{k0}\right| }\times
\right. \right. \\
&&\left. \left. \times \int_{0}^{z}\frac{dz^{\prime }}{\sqrt{\Omega
_{m0}(1+z^{\prime })^{3}+\Omega _{c0}\sqrt{\bar{A}+(1-\bar{A})(1+z^{\prime
})^{6}}+\Omega _{k0}(1+z^{\prime })^{2}}}\right] \right\} \quad ,
\label{mMmu0}
\end{eqnarray}
where $f(x)=\sinh (x)$ for $\Omega _{k0}>0$ (open Universe with $k<0$), $%
f(x)=\sin (x)$ for $\Omega _{k0}<0$ (closed Universe with $k>0$) and $f(x)=x$
for $\Omega _{k0}=0$ (flat Universe with $k=0$).

Following the same lines, we can obtain the expression of the age of the
Universe today:
\begin{equation}
t_{0}=\int_{0}^{\infty }\frac{dz^{\prime }}{H_{0}(1+z^{\prime })\left[
\Omega _{m0}(1+z^{\prime })^{3}+\Omega _{c0}\sqrt{\bar{A}+(1-\bar{A}%
)(1+z)^{6}}+\Omega _{k0}(1+z^{\prime })^{2}\right] }\quad ,
\end{equation}
which will be another parameter evaluated for the different models.

The deceleration parameter $q$ is defined as $q=-1-\dot{H}/H^{2}$, and its
value calculated today, using Eq. (\ref{H2b}), reads
\begin{equation}
q_{0}=\frac{\Omega _{m0}+\Omega _{c0}(1-3\bar{A})}{2},  \label{q0}
\end{equation}
so depending on the values of the three parameters above, an accelerating
Universe ($q_{0}<0$) or decelerating Universe ($q_{0}>0$) is obtained. When $%
\bar{A}=1$ the well-known result for the cosmological constant model is
recovered, $q_{0}=(\Omega _{m0}/2)-\Omega _{c0}$.

Another useful result is the calculation whether the Universe will expand
eternally or not from the present time. A simple method used here searches
for the roots of $a(t)$ in the r.h.s. of Eq. (\ref{H2b}); if there is one
that is a real value and greater than $a_{0}=1$, then the Universe is not
always expanding in the future.

We proceed by fitting the SNe Ia data using the model described above.
Essentially, we compute the quantity distance moduli,
\begin{equation}
\mu _{0}=5\log \left( \frac{D_{L}}{M\!pc}\right) +25\quad ,
\end{equation}
and compare the same distance moduli as obtained from observations. The
quality of the fitting is characterized by the $\chi ^{2}$ parameter of the
least-squares statistic, as defined in Ref. \cite{riess},
\begin{equation}
\chi ^{2}=\sum_{i}\frac{\left( \mu _{0,i}^{o}-\mu _{0,i}^{t}\right) ^{2}}{%
\sigma _{\mu _{0},i}^{2}+\sigma _{mz,i}^{2}}\quad .  \label{Chi2}
\end{equation}
In this expression, $\mu _{0,i}^{o}$ is the measured value, $\mu _{0,i}^{t}$
is the value calculated through the model described above, $\sigma _{\mu
_{0},i}^{2}$ is the measurement error, $\sigma _{mz,i}^{2}$ is the
dispersion in the distance modulus due to the dispersion in galaxy redshift
due to peculiar velocities. This quantity we will taken as
\begin{equation}
\sigma _{mz}=\frac{\partial \log D_{L}}{\partial z}\sigma _{z}\quad ,
\end{equation}
where, following Ref. \cite{riess,wang}, $\sigma _{z}=200\,km/s$.

\begin{table}[t]
\begin{center}
\begin{tabular}{|c|c|c||c|c|c|}
\hline\hline
&  &  &  &  &  \\[-7pt]
\textbf{SNe Ia} & \textbf{z} & \textbf{$\mu_{0}(\sigma_{\mu0})$} & \textbf{%
SNe Ia} & \textbf{z} & \textbf{$\mu_{0}(\sigma_{\mu 0})$} \\%
[2pt] \hline\hline
&  &  &  &  &  \\[-7pt]
\textbf{1992al} & 0.014 & 34.13 (0.14) & \textbf{1992bp} & 0.080 & 37.96
(0.15) \\[2pt] \hline
&  &  &  &  &  \\[-7pt]
\textbf{1992bo} & 0.018 & 34.88 (0.21) & \textbf{1992br} & 0.087 & 38.09
(0.36) \\[2pt] \hline
&  &  &  &  &  \\[-7pt]
\textbf{1992bc} & 0.020 & 34.77 (0.15) & \textbf{1992aq} & 0.111 & 38.33
(0.23) \\[2pt] \hline
&  &  &  &  &  \\[-7pt]
\textbf{1992ag} & 0.026 & 35.53 (0.20) & \textbf{1996J} & 0.30 & 40.99 (0.25)
\\[2pt] \hline
&  &  &  &  &  \\[-7pt]
\textbf{1992P} & 0.026 & 35.59 (0.16) & \textbf{1996K} & 0.38 & 42.21 (0.18)
\\[2pt] \hline
&  &  &  &  &  \\[-7pt]
\textbf{1992bg} & 0.035 & 36.49 (0.21) & \textbf{1996E} & 0.43 & 42.03 (0.22)
\\[2pt] \hline
&  &  &  &  &  \\[-7pt]
\textbf{1992bl} & 0.043 & 36.53 (0.20) & \textbf{1996U} & 0.43 & 42.34 (0.17)
\\[2pt] \hline
&  &  &  &  &  \\[-7pt]
\textbf{1992bh} & 0.045 & 36.87 (0.17) & \textbf{1997cl} & 0.44 & 42.26
(0.16) \\[2pt] \hline
&  &  &  &  &  \\[-7pt]
\textbf{1990af} & 0.050 & 36.67 (0.25) & \textbf{1995K} & 0.48 & 42.49 (0.17)
\\[2pt] \hline
&  &  &  &  &  \\[-7pt]
\textbf{1993ag} & 0.050 & 37.11 (0.19) & \textbf{1997cj} & 0.50 & 42.70
(0.16) \\[2pt] \hline
&  &  &  &  &  \\[-7pt]
\textbf{1993O} & 0.052 & 37.31 (0.14) & \textbf{1996I} & 0.57 & 42.83 (0.21)
\\[2pt] \hline
&  &  &  &  &  \\[-7pt]
\textbf{1992bs} & 0.064 & 37.63 (0.18) & \textbf{1996H} & 0.62 & 43.01 (0.15)
\\[2pt] \hline
&  &  &  &  &  \\[-7pt]
\textbf{1992ae} & 0.075 & 37.77 (0.19) & \textbf{1997ck} & 0.97 & 44.30
(0.19) \\[2pt] \hline\hline
\end{tabular}
\end{center}
\caption{The SNe Ia data of the 26 supernovae used in this article, obtained
by the template fitting method ($\Delta m_{15}(B)$).}
\label{SNIadata}
\end{table}

In table \ref{SNIadata} the data concerning $26$ type Ia supernovae with the
error bar are displayed. These are essentially the type Ia supernovae
employed in the first works on the problem of the acceleration of the
Universe \cite{riess}. We restrict ourselves to this sample (samples with up
to about one hundred type Ia supernovae are now available), since it
contains some of the better studied SNe Ia. It must be stressed that since
one of the goals of this work is to compare competitive models, the choice
of the sample is not so essential, provided it is not too small neither
contains doubtful data.

There are two methods to determine the relationship between the shape of SNe
Ia light curve and its peak luminosity: MLCS (Multicolor Light Curve Shapes)
\cite{Riess1996a} and a template fitting method ($\Delta m_{15}(B)$) \cite
{Hamuy1995}. For both methods, the fit determines the light curve parameters
and their uncertainties. The set of 26 SNe Ia used in this article are
within the expected statistical uncertainties range of the two methods.
Specifically, we use the parameters obtained by the template fitting method
in our analyses.

\begin{table}[!h]
\begin{center}
\begin{tabular}{|c|c|c|}
\hline\hline
&  &  \\[-7pt]
& Cosmological constant & Cosmological constant \\
Parameters & model with $k\leq0$ & model with $k>0$ \\[2pt] \hline
&  &  \\[-7pt]
$\chi _{\nu }^{2}$ & $0.7743$ & $0.7539$ \\[2pt] \hline
&  &  \\[-7pt]
$H_{0}$ & $61.8$ & $62.4$ \\[2pt] \hline
&  &  \\[-7pt]
$\Omega _{k0}$ & $0.00$ & $-0.80$ \\[2pt] \hline
&  &  \\[-7pt]
$\Omega _{m0}$ & $0.24$ & $0.57$ \\[2pt] \hline
&  &  \\[-7pt]
$\Omega _{\Lambda}$ & $0.76$ & $1.23$ \\[2pt] \hline
&  &  \\[-7pt]
$t_{0}$ & $16.5$ & $15.7$ \\[2pt] \hline
&  &  \\[-7pt]
$q_{0}$ & $-0.64$ & $-0.95$ \\[2pt] \hline\hline
\end{tabular}
\end{center}
\caption{The best-fitting parameters, i.e., when $\protect\chi _{\protect\nu
}^{2}$ is minimum, for each type of space section of the cosmological
constant model. $H_{0}$ is given in $km/M\!pc.s$ and $t_{0}$ in $Gy$.}
\label{tableBestFitCC}
\end{table}

\begin{table}[!h]
\begin{center}
\begin{tabular}{|c|c|c|}
\hline\hline
&  &  \\[-7pt]
& Chaplygin gas model & Chaplygin gas model \\
Parameters & with $k\leq0$ & with $k>0$ \\[2pt] \hline
&  &  \\[-7pt]
$\chi _{\nu }^{2}$ & $0.7606$ & $0.7527$ \\[2pt] \hline
&  &  \\[-7pt]
$H_{0}$ & $62.1$ & $62.5$ \\[2pt] \hline
&  &  \\[-7pt]
$\Omega _{k0}$ & $0.00$ & $-0.44$ \\[2pt] \hline
&  &  \\[-7pt]
$\Omega _{m0}$ & $0.00$ & $0.13$ \\[2pt] \hline
&  &  \\[-7pt]
$\Omega _{c0}$ & $1.00$ & $1.31$ \\[2pt] \hline
&  &  \\[-7pt]
$\bar{A}$ & $0.847$ & $0.858$ \\[2pt] \hline
&  &  \\[-7pt]
$t_{0}$ & $15.1$ & $14.6$ \\[2pt] \hline
&  &  \\[-7pt]
$q_{0}$ & $-0.77$ & $-0.97$ \\[2pt] \hline\hline
\end{tabular}
\end{center}
\caption{The best-fitting parameters, i.e., when $\protect\chi _{\protect\nu
}^{2}$ is minimum, for each type of space section of the Chaplygin gas
model. $H_{0}$ is given in $km/M\!pc.s$, $\bar{A}$ in units of $c$ and $%
t_{0} $ in $Gy$.}
\label{tableBestFitCG}
\end{table}

In the tables \ref{tableBestFitCC} and \ref{tableBestFitCG} we evaluate, in
fact, $\chi _{\nu }^{2}$, the estimated errors for degree of freedom, i.e., $%
\chi ^{2}$ divided by $26$, the number of type Ia supernovae chosen in this
article. The values of $\chi _{\nu }^{2}$ for the best-fitting are indeed
small, between $0.75$ and $0.78$, if compared to other SNe Ia analyses \cite
{riess}, due to our choice of SNe Ia that avoided those with large
observational errors. In previous works \cite{JSP,DDCG}, the restricted case
of spatially flat Universe was analysed by means of best-fitting parameter
estimation.

The table \ref{tableBestFitCC} lists the best-fitting parameters of the
cosmological constant model (the limit $\bar{A}\rightarrow 1$ of the
Chaplygin gas model). When the spatial section is open or flat, the
best-fitting favours a flat Universe with $H_{0}=61.8\ km/M\!pc.s$, $\Omega
_{m0}/\Omega _{\Lambda }=0.24/0.76$ and $t_{0}=16.5\ Gy$, approximately the
same result estimated by other research groups \cite{riess,perlmutter}. But
a closed Universe gives a better fitting of the parameters, since $\chi
_{\nu }^{2}$ is smaller, suggesting a different Universe, with positive
curvature and younger. It is clear that small amounts of $\chi _{\nu }^{2}$
lead to totally different parameters, i.e., the parameter estimation is
highly dependent on $\chi _{\nu }^{2}$.

The best-fitting parameters for the Chaplygin gas model are given by table
\ref{tableBestFitCG}. Slightly smaller values of $\chi _{\nu }^{2}$ favour
the Chaplygin gas model over the cosmological constant model, independent of
the spatial section type. The best-of-all fitting suggests a closed Universe
dominated by the Chaplygin gas, featuring a positive curvature, small value
of $\Omega _{m0}$, and the smallest age. In the case of open or flat spatial
section, the best-fit gives a limit case of a flat Universe without cold
dark matter content, just filled by the energy density of the Chaplygin gas.

Nevertheless, the $\chi _{\nu }^{2}$ best-fitting analysis has many
limitations; the following questions cannot be answered. How much worse is
one parameter set compared to other one, for example in table \ref
{tableBestFitCC} or \ref{tableBestFitCG} ? What is the likelihood of a
closed Universe using the Chaplygin gas model (table \ref{tableBestFitCG}) ?
And the likelihood of the age of Universe to be in the range $13$ to $15\ Gy$
?

\section{Bayesian probabilities}

We can answer these questions using the frequentist (traditional) or the
Bayesian probability theory of statistics. See Refs. \cite
{Loredo,Gregory,Drell,Abroe}\ for discussions about them and some
applications in physics. We have chosen the Bayesian approach to avoid some
problems of the frequentist theory: the need of simulating the observational
data, lack of mathematical formalism, etc.

The posterior probability density function (PDF) for the cosmological
parameters $H_{0}$, $\Omega _{m0}$, $\Omega _{c0}$ and $\bar{A}$, given the
set of distance moduli $\mu _{0}$ data and a cosmological model, can be
obtained from the probability of the $\mu _{0}$ data conditional on $%
(H_{0},\Omega _{m0},\Omega _{c0},\bar{A})$, if we use the Bayes's theorem:
\begin{equation}
p(H_{0},\Omega _{m0},\Omega _{c0},\bar{A}\mid \mu _{0})=\frac{p(\mu _{0}\mid
H_{0},\Omega _{m0},\Omega _{c0},\bar{A})\ p(H_{0},\Omega _{m0},\Omega _{c0},%
\bar{A})}{p(\mu _{0})},  \label{BayesTheorem}
\end{equation}
where the prior probabilities $p(H_{0},\Omega _{m0},\Omega _{c0},\bar{A})$
and $p(\mu _{0})$ are considered here to be constants because we have no
prior constraints on the parameters or on the data, besides some forbidden
regions of the parameter space corresponding to unphysical universes. So,
the probability of $(H_{0},\Omega _{m0},\Omega _{c0},\bar{A})$ conditional
on $\mu _{0}$ is proportional to the probability of $\mu _{0}$ conditional
on $(H_{0},\Omega _{m0},\Omega _{c0},\bar{A})$:
\begin{equation}
p(H_{0},\Omega _{m0},\Omega _{c0},\bar{A}\mid \mu _{0})\propto p(\mu
_{0}\mid H_{0},\Omega _{m0},\Omega _{c0},\bar{A}).  \label{ppmu0}
\end{equation}

Each distance moduli $\mu _{0,i}$ is considered as independent with a
Gaussian distribution, then the probability of the set of distance moduli $%
\mu _{0}$ conditional on $(H_{0},\Omega _{m0},\Omega _{c0},\bar{A})$ is the
product of the Gaussians, as explained in Ref. \cite{riess}:
\begin{equation}
p(\mu _{0}\mid H_{0},\Omega _{m0},\Omega _{c0},\bar{A})=\prod\limits_{i}%
\frac{1}{\sqrt{2\pi (\sigma _{\mu _{0},i}^{2}+\sigma _{mz,i}^{2})}}\exp %
\left[ -\frac{\left( \mu _{0,i}^{o}-\mu _{0,i}^{t}\right) ^{2}}{2(\sigma
_{\mu _{0},i}^{2}+\sigma _{mz,i}^{2})}\right] ,  \label{pmu0prod}
\end{equation}
which can be written using the $\chi ^{2}$ calculated previously, Eq. (\ref
{Chi2}), because the product of exponentials is the exponential of the sum
present in $\chi ^{2}$,
\begin{equation}
p(\mu _{0}\mid H_{0},\Omega _{m0},\Omega _{c0},\bar{A})\varpropto \exp
\left( -\frac{\chi ^{2}}{2}\right) .  \label{pmu0}
\end{equation}

Finally, by using the relation (\ref{ppmu0}), the probability of the
parameters $(H_{0},\Omega _{m0},\Omega _{c0},\bar{A})$ conditional on the
set of distance modulus $\mu _{0}$ can be written in a normalized form:
\begin{equation}
p(H_{0},\Omega _{m0},\Omega _{c0},\bar{A}\mid \mu _{0})=\frac{\exp \left( -%
\frac{\chi ^{2}}{2}\right) }{\int\nolimits_{-\infty }^{\infty }\mathrm{d}%
H_{0}\int\nolimits_{-\infty }^{\infty }\mathrm{d}\Omega
_{m0}\int\nolimits_{-\infty }^{\infty }\mathrm{d}\Omega
_{c0}\int\nolimits_{-\infty }^{\infty }\exp \left( -\frac{\chi ^{2}}{2}%
\right) \mathrm{d} \bar{A}},  \label{pNormalized}
\end{equation}
where the integrals are performed on the allowed region of the parameter
space.

Obviously, as the least-squares error $\chi ^{2}$ statistic is minimum, the
probability is maximum. The values of the parameters $(H_{0},\Omega
_{m0},\Omega _{c0},\bar{A})$ of this maximum are the most likely parameter
values, when thought simultaneously.

But when we ask for the best value of one parameter, the $\chi ^{2}$
analysis is not robust. For example, the $\chi ^{2}$ minimum can be located
in a narrow region with small values of $\chi ^{2}$ whereas another larger
region can also have small values of $\chi ^{2}$, and this information is
not taken into account to determine one of the parameters $(H_{0},\Omega
_{m0},\Omega _{c0},\bar{A})$ for the $\chi ^{2}$ minimum.

With the Bayesian probability (\ref{pNormalized}), or likelihood, or PDF,
defined in the four-dimensional parameter space, we can construct a
probability for one parameter by marginalizing, i.e., integrating with
relation to the other parameters, the so-called marginal probability. For
example, the marginal probability for $H_{0}$ is defined as
\begin{equation}
p(H_{0}\mid \mu _{0})=\int\nolimits_{-\infty }^{\infty }\mathrm{d}\Omega
_{m0}\int\nolimits_{-\infty }^{\infty }\mathrm{d}\Omega
_{c0}\int\nolimits_{-\infty }^{\infty }p(H_{0},\Omega _{m0},\Omega _{c0},%
\bar{A}\mid \mu _{0})\mathrm{d}\bar{A},  \label{pH0}
\end{equation}
where again the integration region is restricted to the parameter space with
physical meaning. The peak of this PDF provides the maximum likelihood
estimate of the parameter $H_{0}$, which is usually different from the value
of $H_{0}$ for $\chi ^{2}$ minimum, because generally, $p(H_{0},\Omega
_{m0},\Omega _{c0},\bar{A}\mid \mu _{0})$ is not a four-dimensional Gaussian
PDF (or some distribution alike). This maximum likelihood estimate of $H_{0}$
is a more robust estimate because the rest of the parameter space is the
integrate region for each value of $H_{0}$, so a larger region with high PDF
values (but not the maximum) is not discarded.

The availability of a PDF for a parameter also allows us to calculate the
likelihood of arbitrary hypothesis, for example: $H_{0}$ greater than some
value, $H_{0}$ between some range, etc. They are performed by calculating
the CDF\ (cumulative distribution function), i.e., the integral of the PDF
over the specified region. The inverse problem also appears: for a given
likelihood value, calculate the region in the parameter space which has a
CDF equal to some likelihood value.

A well-known example is the credible or confidence region of $1\sigma $, $%
2\sigma $ and $3\sigma $ likelihoods. To define this type of region, we use
the property that Gaussian PDF function, $\sigma ^{-1}(2\pi )^{-1/2}\exp
(-x^{2}/2\sigma ^{2})$, when integrated over the region $-1\sigma <x<1\sigma
$, gives a probability of approx. $68.27\%$; with the region $-2\sigma
<x<2\sigma $ the obtained likelihood\ is approx. $95.45\%$; and the range $%
-3\sigma <x<3\sigma $ yields a CDF of approx. $99.73\%$. So, $\sigma $
values mean probabilities.

For an arbitrary PDF function $p(x)$, the calculation of the $2\sigma $ (for
example) credible or confidence region in one dimension is usually defined
as follows: obtain the PDF level between zero and the peak PDF (when $%
x=x_{0} $) which have the intersection with $p(x)$, $x_{-}$ and $x_{+}$, so
that $\int\nolimits_{x_{-}}^{x+}p(x)\mathrm{d}x\simeq 95.45$. The estimation
of $x$ using $2\sigma $ credible region is described as $%
(x_{0})_{x_{-}-x_{0}}^{x_{+}-x_{0}}$, meaning that the PDF is peaked at $%
x_{0}$ and the CDF of the region $x_{-}<x<x_{+}$ is equal to $2\sigma $ ($%
95.45\%$). See, for example, the thin line PDF in figure \ref{figOmegam0},
with its $1\sigma $ and $2\sigma $ credible regions, PDF levels and
intersections. In $n$ dimensions, the intersection produces a $n$%
-dimensional region that becomes the integrate region, for an example in two
dimensions, figure \ref{figOmegasCC} shows the $1\sigma $, $2\sigma $ and $%
3\sigma $ credible regions.

The marginal joint probability (PDF) as function of two parameters, for
example $p(\Omega _{m0},\Omega _{c0})$, is an integral in the two
dimensional parameter space of $(H_{0},\bar{A})$,
\begin{equation}
p(\Omega _{m0},\Omega _{c0}\mid \mu _{0})=\int\nolimits_{-\infty }^{\infty
}H_{0}\int\nolimits_{-\infty }^{\infty }p(H_{0},\Omega _{m0},\Omega _{c0},%
\bar{A}\mid \mu _{0})\mathrm{d}\bar{A}.  \label{pOmegam0c0}
\end{equation}
Analogously, the marginalization method can be used to obtain, for example, $%
p(H_{0},\Omega _{m0},\Omega _{c0}\mid \mu _{0})$ from $p(H_{0},\Omega
_{m0},\Omega _{c0},\bar{A}\mid \mu _{0})$ by means of one integral over the $%
\bar{A}$ parameter space (or marginalizing over $\bar{A}$).

Any quantity depending on the parameters can also have a probability for it.
For example, in the case of $t_{0}$, the dynamical age of Universe, the PDF
is obtained from:
\begin{equation}
p(t_{0}\mid \mu _{0})=\int\nolimits_{-\infty }^{\infty }\mathrm{d}%
H_{0}\int\nolimits_{-\infty }^{\infty }\mathrm{d}\Omega
_{m0}\int\nolimits_{-\infty }^{\infty }\mathrm{d}\Omega
_{c0}\int\nolimits_{-\infty }^{\infty }p(H_{0},\Omega _{m0},\Omega _{c0},%
\bar{A}\mid t_{0},\mu _{0})\mathrm{d}\bar{A},  \label{pt0}
\end{equation}
where the integrand gives the likelihood of the parameters which give a
certain age $t_{0}$, so the four-dimensional integral is a sum of the
likelihood of all possible parameters which produces a Universe with age $%
t_{0}$. To avoid the computation of a four-dimensional integral for each
value $t_{0}$, it is better to calculate the age of Universe today for the
four-dimensional parameter space and store the cumulative probabilities for
each value of $t_{0}$.

\section{Analyses of the estimated parameters}

We have performed some long calculations using the Bayesian approach to
obtain the parameter estimations and answers for some hypothesis. In the
Chaplygin gas case, the five parameters, $(H_{0},\Omega _{k0},\Omega
_{m0},\Omega _{c0},\bar{A})$, the age of the Universe $t_{0}$ and the
deceleration parameter $q_{0}$ were estimated with a central value and a $%
2\sigma $ ($95.45\%$) credible region. Each independent parameter estimate
used a marginal likelihood of the type of Eq. (\ref{pH0}), where
three-dimensional integrals are computed for each value of the parameter
(and in the integrand, $\chi ^{2}$ needs the calculation of about one
hundred numerical integrals, or four times the number of supernovae).

An ideal calculation to compute the n-dimensional integrals would include an
infinite number of samples of a parameter space with infinite volume, but in
practical estimations we chose a finite region of the parameter space (such
that outside it the probabilities are almost null) and a finite number of
samples. Considering $n$ samples for each parameter dimension, the
estimation of one parameter needs $n^{4}$ computations of $\chi ^{2}$ (as
there are four independent parameters), and by proper marginalization the
PDF of each parameter is calculated. For this article, $n\approx 30$, and
the number of computations of $\chi ^{2}$ is at least $n^{4}\approx 10^{6}$
just for the Chaplygin gas model (some calculations are repeated because it
is not worth storing $10^{6}$ results due to computational memory
constraints), or $100n^{4}\approx 10^{8}$ numerical integrals of the type in
Eq. (\ref{mMmu0}). We also calculate $n^{4}\approx 10^{6}$ times the age of
Universe today to discard paramater space regions which are not physial ($%
t_{0}$ is not real, etc) and to estimate $t_{0}$. Likewise, the $q_{0}$
estimate demands $n^{4}$ calculations of Eq. (\ref{q0}) for $q_{0}$. The
probability of eternally expanding Universe consists of $n^{4}$ calculations
of the roots of Eq. (\ref{H2b}). The other cases listed in tables \ref
{tableCC} and \ref{tableGC} demands $n^{3}$ or $n^{2}$ calculations, instead
of $n^{4}$.

\begin{table}[!t]
\begin{center}
\begin{tabular}{|c|c|c|c|}
\hline\hline
&  &  &  \\[-7pt]
& Cosmological constant & Cosmological constant & Cosmological constant \\
Parameters & model & model with $k=0$ & model with $\Omega _{m0}=0$ \\%
[2pt] \hline
&  &  &  \\[-7pt]
$H_{0}$ & $62.2\pm 3.1$ & $62.2\pm 3.1$ & $61.4\pm 3.0$ \\[2pt] \hline
&  &  &  \\[-7pt]
$\Omega _{k0}$ & $-0.80_{-1.34}^{+1.45}$ & $0$ & $0.55_{-0.37}^{+0.50}$ \\%
[2pt] \hline
&  &  &  \\[-7pt]
$\Omega _{m0}\geqslant 0$ & $0.58_{-0.58}^{+0.56}$ & $0.24_{-0.16}^{+0.21}$
& $0$ \\[2pt] \hline
&  &  &  \\[-7pt]
$\Omega _{\Lambda }$ & $1.21_{-0.91}^{+0.81}$ & $0.76_{-0.21}^{+0.16}$ & $%
0.45_{-0.50}^{+0.37}$ \\[2pt] \hline
&  &  &  \\[-7pt]
$t_{0}$ & $15.4_{-1.9}^{+3.4}$ & $15.5_{-1.9}^{+3.3}$ & $19.3_{-3.1}^{+5.3}$
\\[2pt] \hline
&  &  &  \\[-7pt]
$q_{0}$ & $-0.99_{-0.52}^{+0.75}$ & $-0.75_{-0.14}^{+0.44}$ & $%
-0.47_{-0.37}^{+0.50}$ \\[2pt] \hline
&  &  &  \\[-7pt]
$p(\Omega _{k0}<0)$ & $81.6\%(1.33\sigma )$ & $-$ & $0.03\%$ \\[2pt] \hline
&  &  &  \\[-7pt]
$p(\Omega _{\Lambda }>0)$ & $99.8\%(3.03\sigma )$ & $99.999993\%(4.86\sigma
) $ & $95.7\%(2.02\sigma )$ \\[2pt] \hline
&  &  &  \\[-7pt]
$p(q_{0}<0)$ & $99.6\%(2.86\sigma )$ & $99.97\%(3.64\sigma )$ & $%
96.1\%(2.06\sigma )$ \\[2pt] \hline
&  &  &  \\[-7pt]
$p(\dot{a}>0)$ & $93.7\%(1.86\sigma )$ & $99.1\%(2.61\sigma )$ & $%
37.8\%(0.49\sigma )$ \\[2pt] \hline\hline
\end{tabular}
\end{center}
\caption{The estimated parameters for the cosmological constant model, using
the Bayesian analysis to obtain the peak of the marginal probability and the
credible region for each parameter. $H_{0}$ is given in $km/M\!pc.s$, $\bar{A%
}$ in units of $c$ and $t_{0}$ in $Gy$.}
\label{tableCC}
\end{table}

\begin{table}[!h]
\begin{center}
\begin{tabular}{|c|c|c|c|c|}
\hline\hline
&  &  &  &  \\[-7pt]
& Chaplygin gas & CGM & CGM & CGM with $k=0$ \\
Parameters & model & with $k=0$ & with $\Omega _{m0}=0$ & and $\Omega
_{m0}=0.04$ \\[2pt] \hline
&  &  &  &  \\[-7pt]
$H_{0}$ & $62.1_{-3.4}^{+3.3}$ & $61.4\pm 2.8$ & $61.9\pm 2.8$ & $61.8\pm
2.8 $ \\[2pt] \hline
&  &  &  &  \\[-7pt]
$\Omega _{k0}$ & $-0.84_{-1.23}^{+1.51}$ & $0$ & $0.17_{-1.58}^{+0.83}$ & 0
\\[2pt] \hline
&  &  &  &  \\[-7pt]
$\Omega _{m0}\geqslant 0$ & $0.00_{-0.00}^{+0.82}$ & $0.00_{-0.00}^{+0.35}$
& $0$ & 0.04 \\[2pt] \hline
&  &  &  &  \\[-7pt]
$\Omega _{c0}\geqslant 0$ & $1.40_{-1.16}^{+1.15}$ & $1.0_{-0.35}^{+0.00}$ &
$0.83_{-0.83}^{+1.59}$ & 0.96 \\[2pt] \hline
&  &  &  &  \\[-7pt]
$0\leqslant \bar{A}\leqslant 1$ & $0.93_{-0.21}^{+0.07}$ & $%
0.93_{-0.20}^{+0.07}$ & $0.78_{-0.19}^{+0.22}$ & $0.87_{-0.18}^{+0.13}$ \\%
[2pt] \hline
&  &  &  &  \\[-7pt]
$t_{0}$ & $14.2_{-1.5}^{+2.8}$ & $13.1_{-1.0}^{+0.5}$ & $14.5_{-1.7}^{+2.9}$
& $14.8_{-1.5}^{+2.4}$ \\[2pt] \hline
&  &  &  &  \\[-7pt]
$q_{0}$ & $-0.98_{-0.62}^{+1.02}$ & $-0.65_{-0.27}^{+0.32}$ & $%
-0.56_{-0.26}^{+0.61}$ & $-0.81_{-0.16}^{+0.28}$ \\[2pt] \hline
&  &  &  &  \\[-7pt]
$p(\Omega _{k0}<0)$ & $84.0\%(1.41\sigma )$ & $-$ & $55.7\%(0.77\sigma )$ & $%
-$ \\[2pt] \hline
&  &  &  &  \\[-7pt]
$p(q_{0}<0)$ & $97.5\%(2.24\sigma )$ & $99.96\%(3.55\sigma )$ & $%
95.2\%(1.98\sigma )$ & $99.997\%(4.17\sigma )$ \\[2pt] \hline
&  &  &  &  \\[-7pt]
$p(\dot{a}>0)$ & $93.9\%(1.87\sigma )$ & $99.7\%(2.99\sigma )$ & $%
63.8\%(0.91\sigma )$ & $100\%$ \\[2pt] \hline\hline
\end{tabular}
\end{center}
\caption{The estimated parameters for the Chaplygin gas model, using the
Bayesian analysis to obtain the peak of the marginal probability and the
credible region for each parameter. $H_{0}$ is given in $km/M\!pc.s$, $\bar{A%
}$ in units of $c$ and $t_{0}$ in $Gy$.}
\label{tableGC}
\end{table}

Tables \ref{tableCC} and \ref{tableGC} summarizes the Bayesian parameter
estimation results for seven different cases. As expected, the central
values of the Bayesian parameter estimation and the best-fitting parameters
of tables \ref{tableBestFitCC} or \ref{tableBestFitCG} are different. In
most cases, the likelihood functions (PDF) of the parameters have a shape
similar to a Gaussian function, but with some asymmetry, consequently they
are not shown here because the central value and credible region description
are enough to realize the PDF behaviour.

\subsection{The Hubble parameter $H_{0}$}

The parameter estimation of $H_{0}$ is almost the same for all cases, around
$62\ km/M\!pc.s$ with a narrow credible region of $\pm 3\ km/M\!\!pc.s$.
This estimation is compatible with other SNe Ia analyses \cite
{riess,perlmutter}. Hence, the choice of Chaplygin gas model or $\Lambda CDM$
does not have any significant consequence on the $H_{0}$ estimates we have
made.

\subsection{The curvature density parameter $\Omega _{k0}$}

Likewise, the curvature density parameter $\Omega _{k0}$ estimation is
almost the same, $-0.84_{-1.23}^{+1.51}$ and $-0.80_{-1.34}^{+1.45}$,
assuming the Chaplygin gas model or $\Lambda CDM$, respectively, and in both
cases a closed Universe ($\Omega _{k0}<0$) is preferred at $84.0\%$ ($%
1.41\sigma $) and $81.6\%$ ($1.33\sigma $) confidence levels, also
respectively, in agreement with the conclusions of many other previous works
on the subject \cite{riess,perlmutter}. In the $CGM$, a spatially flat
Universe is ruled out at $67.5\%$ ($0.98\sigma $) confidence level, i.e.,
the PDF of $\Omega _{k0}=0$ is smaller than the PDF of $-1.42<\Omega _{k0}<0$
and this region has a CDF of $67.5\%$ ($0.98\sigma $). The $\Lambda CDM$
model shows a similar behaviour, a spatially flat Universe is ruled out at $%
63.3\%$ ($0.90\sigma $) level, i.e., the PDF of $\Omega _{k0}=0$ is smaller
than the PDF of $-1.46<\Omega _{k0}<0$.

But assuming the hypothesis that $\Omega _{m0}=0$, describing a Universe
empty of cold dark matter and totally dominated by the Chaplygin gas, we
have quite different estimates. For the $CGM$, the Bayesian analysis gives $%
\Omega _{k0}=0.17_{-1.58}^{+0.83}$, with a closed Universe favoured at $%
55.7\%$ ($0.77\sigma $), and a flat Universe is strongly preferred at $%
83.3\% $ ($1.38\sigma $) confidence level, i.e., the PDF of $\Omega _{k0}=0$
is smaller than the PDF of $0<\Omega _{k0}<0.30$ and this region has a small
CDF of $16.7\%$.

The same hypothesis, $\Omega _{m0}=0$, applied to the $\Lambda CDM$ model,
leads to the estimation $\Omega _{k0}=0.55_{-0.37}^{+0.50}$, so now an open
Universe is quite favourable at $99.97\%$ ($3.60\sigma $) confidence level.
A spatially flat Universe is ruled out at $99.92\%$ ($3.36\sigma $)
confidence level, i.e., the PDF of $\Omega _{k0}=0$ is smaller than the PDF
of $0<\Omega _{k0}<1.43$.

Therefore, the $\Omega _{m0}=0$ case clearly discriminates the Chaplygin gas
and the $\Lambda CDM$ models, as their estimates are quite different. This
discrimination is still enhanced if the additional hypothesis of spatially
flat Universe is assumed (to be compatible with inflationary models of the
primordial Universe, or some CMB estimations, for example), when the $CGM$
agrees at a high confidence level while $\Lambda CDM$ is ruled out.

\subsection{The cold dark matter density parameter $\Omega _{m0}$}

\begin{figure}[!t]
\begin{center}
\includegraphics[scale=0.80]{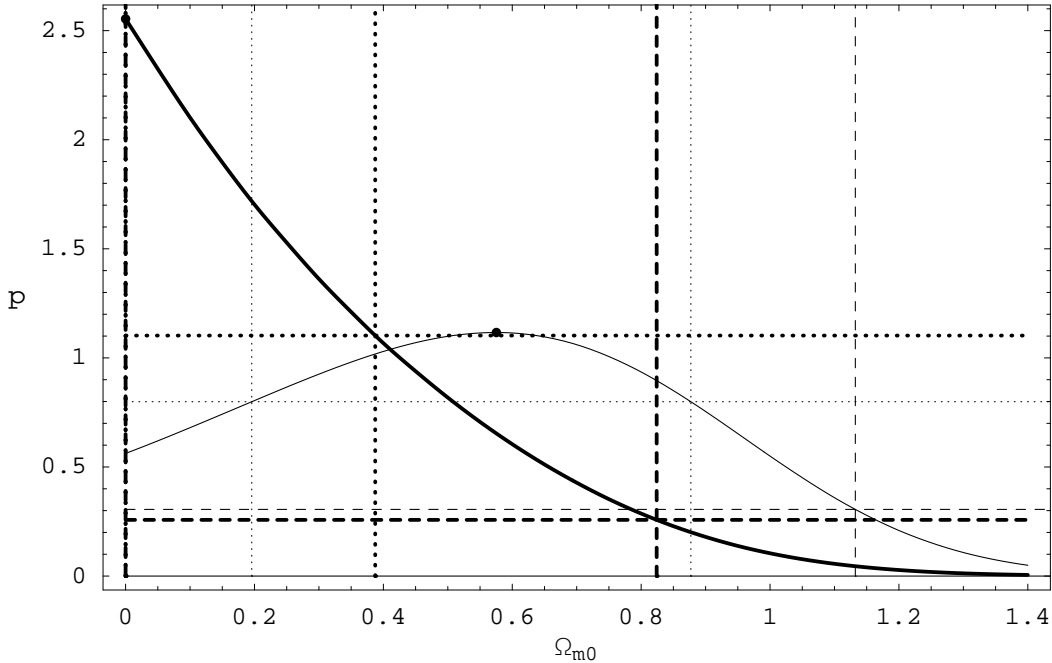}
\end{center}
\caption{{\protect\footnotesize The PDF of $\Omega _{m0}$ for the Chaplygin
gas model (bold lines) and the cosmological constant model (thin lines). The
solid lines are the PDF, the $2\protect\sigma $ ($95.45\%$) credible regions
are given by dashed lines and the $1\protect\sigma $ ($68.27\%$) regions are
delimited by dotted lines. The Chaplygin gas case has maximum probability $%
2.55$ for $\Omega _{m0}=0.0$, the $2\protect\sigma $ credible region reads $%
0\leqslant \Omega _{m0}<0.82$ with probability level $0.26$, and the $1%
\protect\sigma $ credible region is given by $0\leqslant \Omega _{m0}<0.39$
with PDF level $1.10$. The $\Lambda CDM$ has probability peak of $1.11$ when
$\Omega _{m0}=0.58$, the $2\protect\sigma $ credible region reads $%
0\leqslant \Omega _{m0}<1.13$ with probability level $0.31$, and the $1%
\protect\sigma $ credible region is given by $0.20\leqslant \Omega
_{m0}<0.88 $ with PDF level $0.80$. This graphics clearly shows that small
values of $\Omega _{m0}$ are more preferred by the Chaplygin gas model. }}
\label{figOmegam0}
\end{figure}

But the cold dark matter density parameter $\Omega _{m0}$ depends quite on
the cosmological models. Figure \ref{figOmegam0} clearly shows that the PDF
for the Chaplygin gas and the $\Lambda CDM$\ models behave differently, $%
\Omega _{m0}=0.00_{-0.00}^{+0.82}$ and $\Omega _{m0}=0.58_{-0.58}^{+0.56}$,
respectively. So the Chaplygin gas model favours small values of $\Omega
_{m0}$ at a high level of confidence, for example, $\Omega _{m0}<0.39$ at $%
1\sigma $ ($68.27\%$) confidence level, while $\Lambda CDM$ estimates $%
\Omega _{m0}<0.39$ with a probability of $31.1\%$ ($0.40\sigma $), or, $%
\Omega _{m0}=0.58_{-0.28}^{+0.30}$ at $1\sigma $ level.

While the most favoured value of $\Omega _{m0}$ is zero for the $CGM$, the
cosmological constant model rules out $\Omega _{m0}=0$ at $89.5\%$ ($%
1.62\sigma $) confidence level, i.e., the PDF of $\Omega _{m0}=0$ (equal to $%
0.56$) is smaller than the PDF of $0<\Omega _{m0}<0.99$ which has a CDF of $%
89.5\%$.

On the hypothesis that the Universe is spatially flat, the $CGM$ and $%
\Lambda CDM$ have narrower confidence regions, see tables \ref
{tableBestFitCC} and \ref{tableBestFitCG}. It is worth noting that the $CGM$
continues to indicate a $\Omega _{m0}$ peaked at zero, but the $\Lambda CDM$
model rules out $\Omega _{m0}=0$ completly, at $99.94\%$ ($3.44\sigma $)
confidence level (when $0<\Omega _{m0}<0.63$ the PDF is greater than the PDF
of $\Omega _{m0}=0$).

Based on the SNe Ia data, we can conclude that, assuming the Chaplygin gas
model, the estimated values of $\Omega _{m0}$ are decreased (with respect to
$\Lambda CDM$), or if independent estimations of $\Omega _{m0}$ suggest low
values then the Chaplygin gas model is favoured over the $\Lambda CDM$ model.

\subsection{The Chaplygin gas density parameter $\Omega _{c0}$}

The Chaplygin gas density parameter $\Omega _{c0}$ estimation is quite
spread, i.e., it has a large credible region, $\Omega
_{c0}=1.40_{-1.16}^{+1.15}$. Only for the particular case of a spatially
flat Universe ($k=0$) the estimation of $\Omega _{c0}$ is less spread. The $%
\Omega _{m0}=0$ case gives a lower estimate but still strongly spread: $%
\Omega _{c0}=0.83_{-0.83}^{+1.59}$.

For the $\Lambda CDM$ model obtained from the Chaplygin gas model as $\bar{A}%
=1$, the dark energy density $\Omega _{\Lambda }$ is positive at a $%
3.3\sigma $ ($99.8\%$) confidence level, and the estimation $\Omega
_{\Lambda }=1.21_{-0.91}^{+0.81}$ becomes more peaked for a spatially flat
Universe, $\Omega _{\Lambda }=0.76_{-0.21}^{+0.16}$. See table \ref
{tableBestFitCC} for more estimates.

\subsection{$\bar{A}$, the sound velocity of the Chaplygin gas}

\begin{figure}[t]
\begin{center}
\includegraphics[scale=0.80]{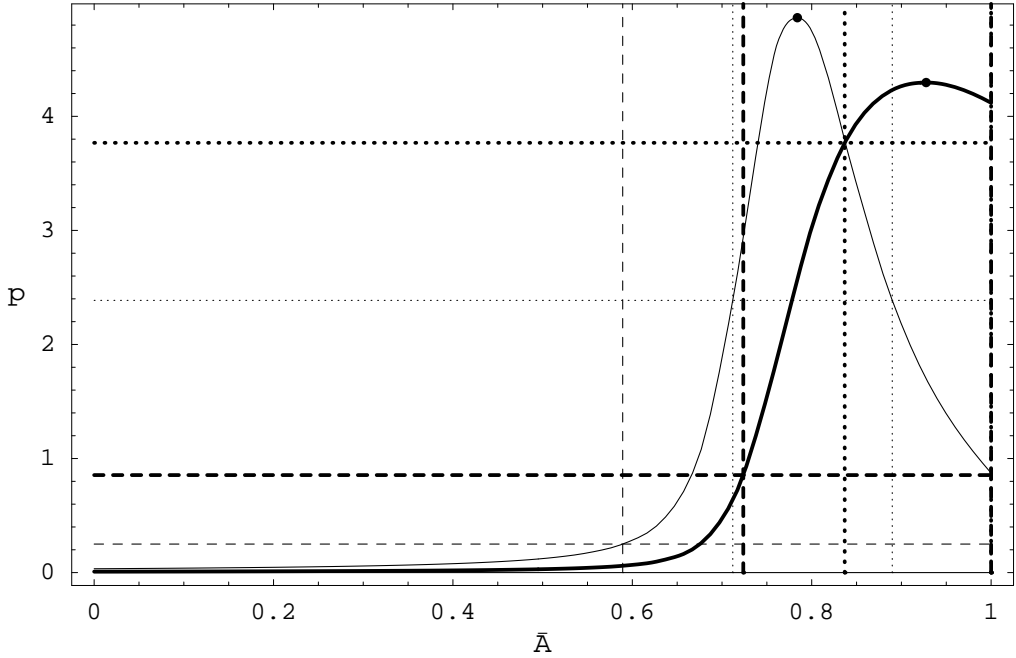}
\end{center}
\caption{{\protect\footnotesize The PDF of $\bar{A}$ for the Chaplygin gas
model (bold lines) and the same with $\Omega _{m0}=0$ (thin lines). The
solid lines are the PDF, the $2\protect\sigma $ ($95.45\%$) credible regions
are given by dashed lines and the $1\protect\sigma $ ($68.27\%$) regions are
delimited by dotted lines. The non-restricted case has maximum probability $%
4.30$ for $\bar{A}=0.93$, the $2\protect\sigma $ credible region reads $%
0.72\leqslant \bar{A}<1$ with probability level $0.85$, and the $1\protect%
\sigma $ credible region is given by $0.84\leqslant \bar{A}<1$ with PDF
level $3.77$. The probability of the $\Lambda CDM$ obtained when $\bar{A}=1$
is $4.12$, so the Chaplygin gas model is more likely (has greater PDF) in
the range $0.87\leqslant \bar{A}<1$, which has a CDF of $55.3\%$ ($0.76%
\protect\sigma $). The $\Omega _{m0}=0$ case has probability peak of $4.86$
when $\bar{A}=0.78$, the $2\protect\sigma $ credible region reads $%
0.59\leqslant \bar{A}<1$ with probability level $0.25$, the $1\protect\sigma
$ credible region is given by $0.71\leqslant \bar{A}<0.89$ with PDF level $%
2.39$, and the $\Lambda CDM$ has a\ PDF level of only $0.87$ which suggests
that the Chaplygin gas model is strongly preferred if there is no cold dark
matter ($\Omega _{m0}=0$) content in the Universe, because the range $%
0.67\leqslant \bar{A}<1$ has a CDF of $91.8\%$ ($1.74\protect\sigma $) such
that the PDF level is greater than $0.87$. }}
\label{figA}
\end{figure}

The analysis of the parameter $\bar{A}$ is very conclusive, because $\Lambda
CDM$ is a special case of the Chaplygin gas model when $\bar{A}=1$,
therefore we can compare the probabilities of each cosmological model.
Indeed, figure \ref{figA} shows that the best estimation reads $\bar{A}%
=0.93_{-0.21}^{+0.07}$, with peak probability level $4.30,$ while the $%
\Lambda CDM$ limit ($\bar{A}=1$) has a slightly lower likelihood of $4.12$.
More specifically, the Chaplygin gas model is the more preferred model over
the range $0.87\leqslant \bar{A}<1$ because the PDF is greater than $4.12$,
and this region represents $55.3\%$ ($0.76\sigma $) of the total
probability. So, even if there is no additional hypothesis about the
curvature parameter $\Omega _{k0}$ or the cold dark matter density parameter
$\Omega _{m0}$, the Bayesian analysis explicitly estimates that the
Chaplygin gas model is more likely than $\Lambda CDM$ at $0.76\sigma $ ($%
55.3\%$) confidence level, although the peak likelihood is just $4\%$
greater than $\Lambda CDM$ likelihood.

Under the hypothesis that $\Omega _{m0}=0$, so the Chaplygin gas is the only
content of the Universe, figure \ref{figA} gives $\bar{A}%
=0.78_{-0.19}^{+0.22}$ and indicates that the Chaplygin gas model has a peak
likelihood of $4.86$, quite greater than the $\Lambda CDM$ model probability
level of $0.87$ (for $\bar{A}=1$), and on a broader region $0.67\leqslant
\bar{A}<1$ the PDF is greater than $0.87$. The parameter $\bar{A}$ is inside
this region with likelihood of $91.8\%$ ($1.74\sigma $), i.e., the Chaplygin
gas model is preferred over the $\Lambda CDM$ model at $1.74\sigma $ ($%
91.8\% $) confidence level.

Assuming now the hypothesis of a flat Universe, $\Omega _{k0}=0$, we obtain $%
\bar{A}=0.93_{-0.20}^{+0.07}$ with a peak likelihood of $5.12$, greater than
the $\Lambda CDM$ model probability level of $4.18$ (for $\bar{A}=1$). The
region $0.87\leqslant \bar{A}<1$ has a PDF greater than $4.18$, which has a
CDF of $63.5\%$ ($0.91\sigma $), i.e., the Chaplygin gas model is preferred
over $\Lambda CDM$ at $63.5\%$ ($0.91\sigma $) confidence level.

In the framework of the cosmological model estimations based on SNe Ia, the
baryonic matter has the same behaviour of the cold dark matter, so we can
assume a typical value of $0.04$ for the baryonic density parameter by
setting $\Omega _{m0}=0.04$. Still imposing the hypothesis of a flat
Universe, the last column of table \ref{tableBestFitCG} shows that $\bar{A}%
=0.87_{-0.18}^{+0.13}$ with a maximum PDF of $5.56$ and the region $%
0.59\leqslant \bar{A}<1$ having greater PDF than the PDF of $\Lambda CDM$
limit ($\bar{A}=1$), $0.13$. The CDF of this region means that the $CGM$ is
preferred over $\Lambda CDM$ at $99.4\%$ ($2.72\sigma $) confidence level,
under the hypothesis $\Omega _{k0}=0$ and $\Omega _{m0}=0.04$. Not shown in
\ref{tableBestFitCC}, the $\Lambda CDM$ model parameter estimates read, $%
H_{0}=64.4_{-2.1}^{+2.2}\,km/M\!pc.s$ and $t_{0}=24.2_{-0.8}^{+0.8}\,Gy$,
with an age of the Universe excessively higher than those obtained through
other astronomical estimations.

Combining both hypothesis, $\Omega _{k0}=0$ and $\Omega _{m0}=0$, the
parameter estimation is quite physically acceptable for the $CGM$. Not show
explicitly in table \ref{tableBestFitCG}, we have obtained $%
H_{0}=61.9_{-2.8}^{+2.8}\,km/M\!pc.s$, $\bar{A}=0.85_{-0.18}^{+0.15}\,$, $%
t_{0}=14.7_{-1.5}^{+2.7}\,Gy$ and $q_{0}=-0.81_{-0.13}^{+0.32}$ with $%
p(q_{0}<0)=99.996\%(4.12\sigma )$. The Chaplygin gas model is strongly
favoured over the $\Lambda CDM$ model at $99.93\%$ ($3.38\sigma $)
confidence level, as the PDF of $\bar{A}=0.85$ is equal to $5.53$ and the
likelihood of $\bar{A}=1$ is $0.02$, so a large region $0.47\leqslant \bar{A}%
<1$ has PDF levels greater than $0.02$. Moreover, the $\Lambda CDM$ model
produces totally unphysical estimates: $t_{0}=368_{-12}^{+13}\times 10^{3}Gy$%
.

\subsection{The present age $t_{0}$ of the Universe}

Looking at the dynamical age of the Universe today, $t_{0}$, given in tables
\ref{tableCC} and \ref{tableGC}, we verify that the Chaplygin gas model
estimates lower values of $t_{0}$ as well as narrower credible regions ($%
t_{0}=14.2_{-1.5}^{+2.8}$), when compared to the estimations of the $\Lambda
CDM$ model ($t_{0}=15.4_{-1.9}^{+3.4}$). By means of independent estimations
of $t_{0}$ from different observations (globular agglomerates, etc), the
Chaplygin gas model would be more preferred if, for example, these
estimations give $t_{0}$ between $13Gy$ to $14Gy$.

If we impose some constraints, $k=0$ or $\Omega _{m0}=0$, we still have
lower estimates for $t_{0}$ of the $CGM$ than the $\Lambda CDM$ model
estimations. The latter are strongly sensible to the value of $\Omega _{m0}$%
: for example, if $k=0$, we obtain $t_{0}=24.2_{-0.8}^{+0.8}\,Gy$ for $%
\Omega _{m0}=0.04$ and totally unphysical value of $t_{0}=368_{-12}^{+13}%
\times 10^{3}Gy$ for $\Omega _{m0}=0$. Quite different, the $CGM$ produces $%
t_{0}$ values with a weak dependence on $\Omega _{k0}$ and $\Omega _{m0}$.

\subsection{The deceleration parameter $q_{0}$ of the Universe}

All the estimations based on the Chaplygin gas model or the cosmological
constant model suggest an accelerating Universe today, i.e., $q_{0}<0$, with
a high confidence level (at least more than $95\%$).

\subsection{The eternally expanding Universe}

The probability of having a eternally expanding Universe, given here by
demanding $\dot{a}>0$ during the future evolution of the Universe, is
estimated to be almost one (at least more than $93\%$), excepting in the $%
\Omega _{m0}=0$ case for both the $CGM$ and $\Lambda CDM$ when $p(\dot{a}>0)$
is far from unity, see tables \ref{tableCC} and \ref{tableGC}.

\subsection{Two-dimensional analysis in the $(\Omega _{m0},\Omega
_{_{\Lambda }})$ and $(\Omega _{m0},\Omega _{c0})$ parameter space}

\begin{figure}[!t]
\begin{center}
\includegraphics[scale=0.80]{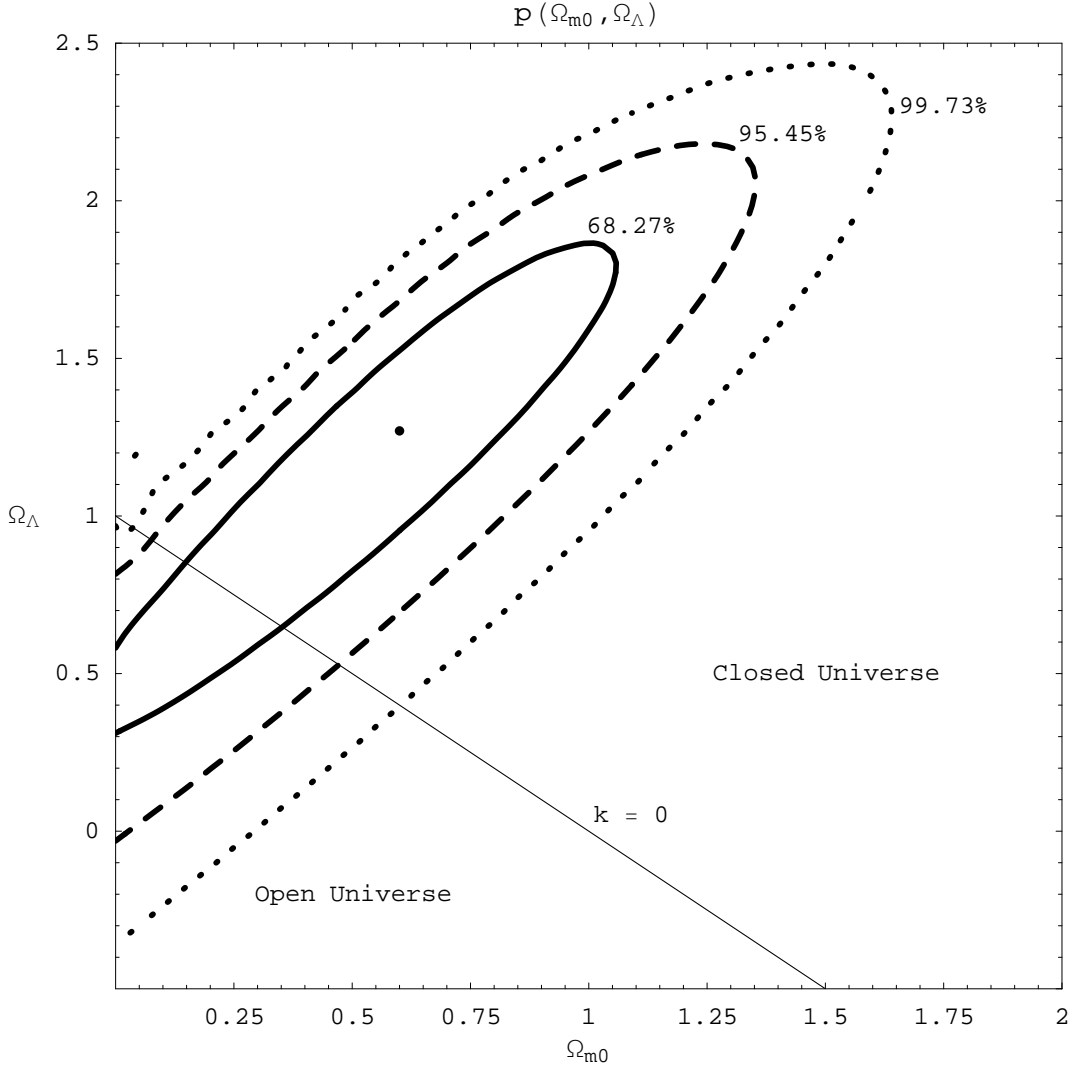}
\end{center}
\caption{{\protect\footnotesize The graphics of the joint PDF as function of
$(\Omega _{m0},\Omega _{\Lambda})$ for the $\Lambda CDM$, where $p(\Omega
_{m0},\Omega _{\Lambda})$ is a integral of $p(H_{0},\Omega _{m0},\Omega
_{\Lambda})$ over the $H_{0}$ parameter space. It shows the well-known shape
of the confidence regions for the cosmological constant model. The joint PDF
peak has the value $2.07$ for $(\Omega _{m0},\Omega_{\Lambda})=(0.60,1.27)$
(shown by the large dot), the $1\protect\sigma $ ($68,27\%$), $2\protect%
\sigma $ ($95,45\%$) and $3\protect\sigma $ ($99,73\%$) credible regions
have joint PDF levels of $0.85$, $0.15$\ and $0.01$, respectively. As $%
\Omega _{k0}+\Omega _{m0}+\Omega _{\Lambda}=1$, the probability for a
spatially flat Universe is on the line $\Omega _{m0}+\Omega _{\Lambda}=1$,
above it we have the region for a closed Universe ($k>0$, $\Omega _{k0}<0$),
and below, the region for an open Universe ($k<0$, $\Omega _{k0}>0$), so the
integral of the PDF over the upper region gives the likelihood of a closed
Universe being $81.6\%$. }}
\label{figOmegasCC}
\end{figure}

\begin{figure}[!t]
\begin{center}
\includegraphics[scale=0.80]{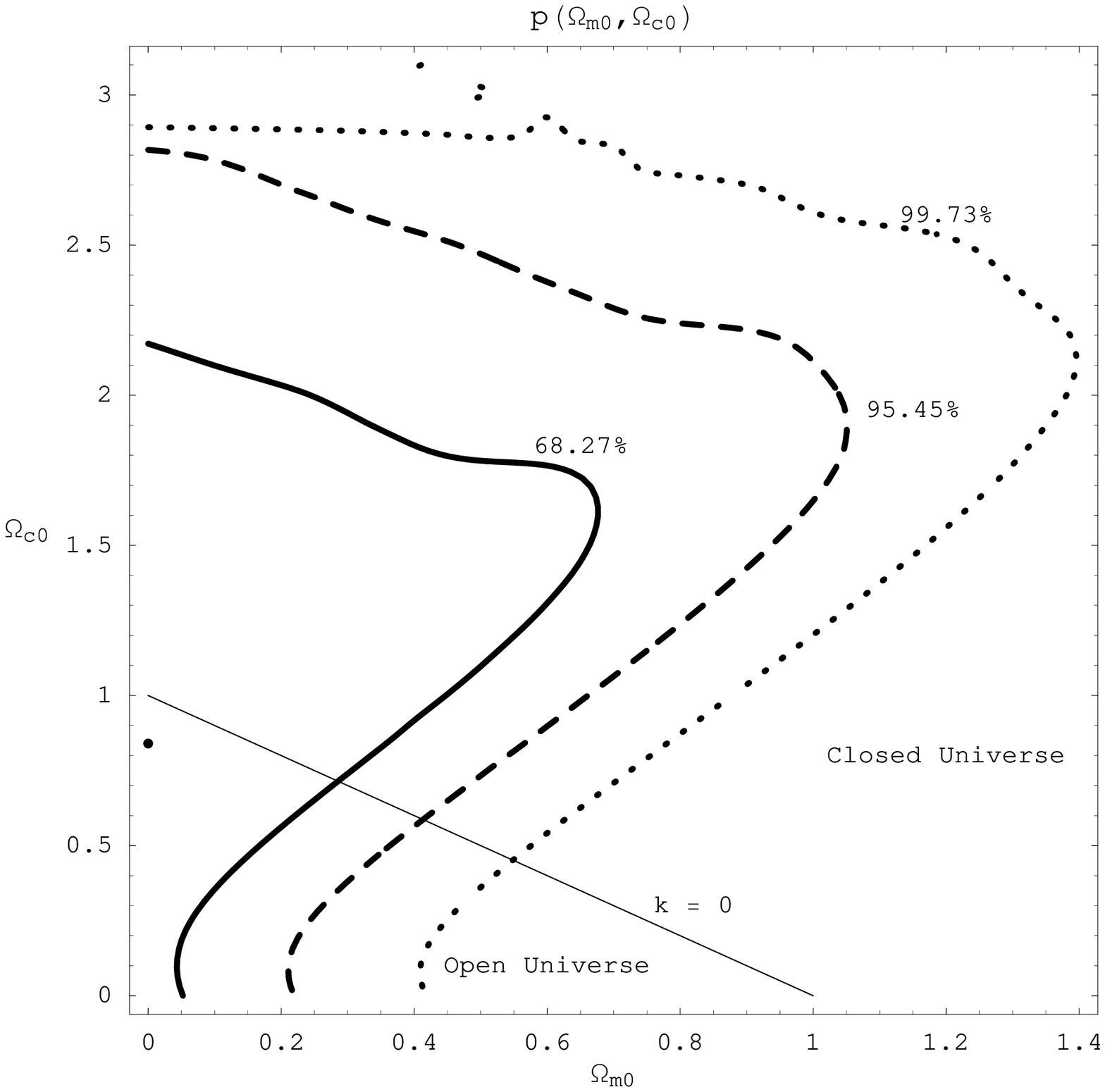}
\end{center}
\caption{{\protect\footnotesize The graphics of the joint PDF as function of
$(\Omega _{m0},\Omega _{c0})$ for the Chaplygin gas model, where $p(\Omega
_{m0},\Omega _{c0})$ is a integral of $p(H_{0},\Omega _{m0},\Omega _{c0},%
\bar{A})$ over the $(H_{0},\bar{A})$ parameter space. It shows a different
shape for the confidence regions with respect to the cosmological constant
model. The joint PDF peak has the value $1.47$ for $(\Omega
_{m0},\Omega_{c0})=(0.00,0.84)$ (shown by the large dot), the $1\protect%
\sigma $ ($68,27\%$), $2\protect\sigma $ ($95,45\%$) and $3\protect\sigma $ (%
$99,73\%$) credible regions have PDF levels of $0.53$, $0.10$\ and $0.01$,
respectively. As $\Omega _{k0}+\Omega _{m0}+\Omega _{c0}=1$, the probability
for a spatially flat Universe is on the line $\Omega _{m0}+\Omega _{c0}=1$,
above it we have the region for a closed Universe ($k>0$, $\Omega _{k0}<0$),
and below, the region for an open Universe ($k<0$, $\Omega _{k0}>0$). The
integral of the joint PDF over the upper region gives the likelihood of a
closed Universe being $84.0\%$, althought the joint PDF peak is located in
the open Universe region. }}
\label{figOmegasGC}
\end{figure}

Finally, it is useful to analyze the two-dimensional joint PDF, for example,
in the parameter space of $(\Omega _{m0},\Omega _{_{\Lambda }})$, where the
PDF $p(\Omega _{m0},\Omega _{\Lambda }\mid \mu _{0})$ is given by the
integral of $p(H_{0},\Omega _{m0},\Omega _{\Lambda }\mid \mu _{0})$ over the
$H_{0}$ space. The well-known credible regions for the $\Lambda CDM$ model
are reproduced here, see figure \ref{figOmegasCC}. It is worth noting that
the probability peak gives $(\Omega _{m0},\Omega _{\Lambda })=(0.60,1.27)$
which are different from the central values listed in table \ref{tableCC}, $%
\Omega _{m0}=0.58$ and $\Omega _{\Lambda }=1.21$, due to the integration
over the $\Omega _{\Lambda }$ space to obtain $p(\Omega _{m0}\mid \mu _{0})$
from $p(\Omega _{m0},\Omega _{\Lambda }\mid \mu _{0})$, and analogously for $%
p(\Omega _{\Lambda }\mid \mu _{0})$. The interpretation is the following : $%
(\Omega _{m0},\Omega _{\Lambda })=(0.60,1.27)$ means that these values are
the most likely simultaneously values of $(\Omega _{m0},\Omega _{\Lambda })$
independent of the parameter $H_{0}$; while $\Omega _{m0}=0.58$ is the most
likely value of $\Omega _{m0}$ independent of the other parameters $%
(H_{0},\Omega _{\Lambda })$, and $\Omega _{\Lambda }=1.21$ is the most
likely value independent of the other parameters $(H_{0},\Omega _{m0})$.
Figure \ref{figOmegam0} shows $p(\Omega _{m0}\mid \mu _{0})$ which is
obtained by integrating $p(\Omega _{m0},\Omega _{\Lambda }\mid \mu _{0})$
over the $\Omega _{\Lambda }$ space.

The contour curves are defined such that the enclosed regions have a
cumulative probability equal to $1\sigma $ ($68,27\%$), $2\sigma $ ($95,45\%$%
) and $3\sigma $ ($99,73\%$) credible levels. For example, this means simply
that $\Omega _{m0}$ and $\Omega _{\Lambda }$ have values simultaneously
inside the $2\sigma $ ($95,45\%$) region with a likelihood of $95,45\%$. The
line $\Omega _{m0}+\Omega _{\Lambda }=1$ represents a spatially flat
Universe ($k=0$, $\Omega _{k0}=0$), the parameter space above corresponds to
a closed Universe ($k>0$, $\Omega _{k0}<0$), and below, to an open Universe (%
$k<0$, $\Omega _{k0}>0$). As the closed Universe region clearly dominates
and presents high PDF values, a closed Universe is estimated at $1.33\sigma $
($81.6\%$) confidence level.

For the case of the Chaplygin gas model, the PDF $p(\Omega _{m0},\Omega
_{c0}\mid \mu _{0})$ is a two-dimensional integral given by Eq. (\ref
{pOmegam0c0}), and figure \ref{figOmegasGC} displays its behaviour with
credible regions quite different from the $\Lambda CDM$. The PDF peak is now
located at $(\Omega _{m0},\Omega _{c0})=(0.0,0.85)$, contrasting with the
central values $\Omega _{m0}=0.00$ and $\Omega _{c0}=1.40$ of table \ref
{tableGC}, due to the Bayesian integrations. Of course, the integral of $%
p(\Omega _{m0},\Omega _{c0}\mid \mu _{0})$ over the $\Omega _{c0}$ space
yields $p(\Omega _{m0}\mid \mu _{0})$, shown in figure \ref{figOmegam0}.
Once more, a spatially closed Universe is strongly favoured at $1.41\sigma $
($84.0\%$) confidence level, despite the PDF peak being inside the spatially
open Universe region.

These two graphics also emphasize the totally different likelihoods for $%
(\Omega _{k0},\Omega _{m0})=(0,0.04)$ in the $CGM$ and $\Lambda CDM$ model.
This point corresponds to $(\Omega _{m0},\Omega _{\Lambda })=(0.04,0.96)$ in
figure \ref{figOmegasCC} for $\Lambda CDM$, with a PDF level of $0.02$, well
below the PDF peak $2.07$, so a large region of the $(\Omega _{m0},\Omega
_{\Lambda })$ parameter space has greater PDF than $0.02$ yielding a CDF of $%
99.6\%$ ($2.91\sigma $), i.e., the simultaneously hypothesis of flat
Universe and $\Omega _{m0}=0.04$ (representing a typical baryonic density
parameter of $0.04$) is ruled out at $99.6\%$ ($2.91\sigma $) assuming the $%
\Lambda CDM$. The limit case of $(\Omega _{k0},\Omega _{m0})=(0,0)$ gives a
confidence level of $99.93\%$ ($3.39\sigma $).

Figure \ref{figOmegasGC} clearly shows that the point $(\Omega _{m0},\Omega
_{c0})=(0.04,0.96)$ is near the maxima of PDF, their PDF levels are $1.41$
and $1.47$, respectively. Therefore, the region with smaller PDF than $1.41$
is almost the whole of the parameter space, and its CDF is equal to $98.6\%$%
, i.e, for the $CDM$ the simultaneously hypothesis of flat Universe and $%
\Omega _{m0}=0.04$ is favoured at $98.6\%$ ($2.45\sigma $) confidence level.
In the same way, the case of $(\Omega _{k0},\Omega _{m0})=(0,0)$ is favoured
at a confidence level of $99.3\%$ ($2.69\sigma $).

\section{Conclusion}

In the present paper constraints on a Chaplygin gas model using type Ia
supernovae data were settled out using a Bayesian statistics. The model
contains as matter content the Chaplygin gas and a pressureless fluid, whose
density parameters are $\Omega _{c0}$ and $\Omega _{m0}$ respectively. The
curvature term $\Omega _{k0}$, the sound velocity of the Chaplygin gas $\bar{%
A}$ and the Hubble parameter $H_{0}$ are also free parameters. At the $%
2\sigma $ level, the results indicate $H_{0}=62.1_{-3.4}^{+3.3}\,km/M\!pc.s$%
, $\Omega _{k0}=-0.84_{-1.23}^{+1.51}$, $\Omega _{m0}=0.0_{-0.0}^{+0.82}$, $%
\Omega _{c0}=1.40_{-1.16}^{+1.15}$, $\bar{A}=c_{s}^{2}=0.93_{-0.21}^{+0.07}%
\,c$, $t_{0}=14.2_{-1.3}^{+2.8}\,Gy$ and $q_{0}=-0.98_{-0.62}^{+1.02}$. This
results must be compared with those obtained replacing the Chaplygin gas by
the cosmological constant ($\bar{A}=1$), for which $H_{0}=62.2\pm
3.1\,km/M\!pc.s$, $\Omega _{k0}=-0.80_{-1.34}^{+1.45}$, $\Omega
_{m0}=0.58_{-0.58}^{+0.56}$, $\Omega _{\Lambda }=1.21_{-0.91}^{+0.81}$, $%
t_{0}=15.4_{-1.9}^{+3.4}\,Gy$ and $q_{0}=-0.99_{-0.52}^{+0.75}$.

The $CGM$ is more likely than the $\Lambda CDM$ model with a confidence
level of $55.3\%$ when all free parameters are considered. This percentage
is much higher if some restriction on the parameters are imposed: fixing $%
\Omega _{k0}=0$, the $CGM$ becomes preferred with respect to $\Lambda CDM$
at $63.5\%$; this confidence level mounts to $91.8\%$ if, instead, $\Omega
_{m0}=0$. The Bayesian analysis of the flat case with Chaplygin gas and
baryonic matter ($\Omega _{b0}=0.04$) implies that the $CGM$ is preferred
with respect to the $\Lambda CDM$ model at $99.4\%$ confidence level.\ In
general, the predicted value for the sound velocity of the Chaplygin gas is
close to the cosmological constant value ($\bar{A}=1$), but after
integrating on the various parameters the $\Lambda CDM$ case becomes quite
disfavoured.

Other important differences between the $CGM$ and $\Lambda CDM$ concern the
pressureless matter parameter and the age of the Universe. For the former,
the $CGM$ favours a zero value for this cold dark matter component, in
agreement with the idea that the Chaplygin gas may unify dark matter and
dark energy, and in contrast with $\Lambda CDM$ where a non-negligible
fraction of the matter in the Universe must appear under the form of dark
matter.

A very remarkable discrimination between the $CGM$ and $\Lambda CDM$ occurs
when the pressureless matter parameter represents the baryonic matter, $%
\Omega _{b0}=0.04$ and the Universe is spatially flat. This case is favoured
at $98.6\%$ of confidence level for the $CGM$, while for the $\Lambda CDM$
model it is excluded with $99.6\%$ of confidence level. This result renders
the $CGM$ quite attractive in view of the predictions of almost all
primordial inflationary scenarios, which lead to a flat Universe, and also
in view of the unification program for dark energy and dark matter through
the Chaplygin gas.

One of the most important conclusions of this work is that the predicted
value for the dark matter parameter $\Omega _{m0}$ is peaked in the zero
value. This reinforces the idea that the Chaplygin gas may unify dark matter
and dark energy as its behaviour in terms of the scale factor suggests. This
unification program has been criticized \cite{ioav,bean} because, besides
some other reasons, the matter power spectrum in a pure $CGM$ exhibits
oscillations that are not observed in the recent $2dFGRS$ data, and the
phase space of possible configuration is highly concentrated around the
cosmological constant value. This seems to be a strong argument against the
unified model. However, we must remark that the authors of Ref. \cite{ioav}
employ an one fluid model. Even if the $CGM$ may unify dark matter and dark
energy, baryons exists anyway, even if in a small fraction ($\Omega
_{b0}\sim 0.04$). This may seems to be irrelevant, but the behaviour of a
two fluid models is generally very different from an one fluid model (see,
for example, Ref. \cite{glauber}), and this point deserves, in our opinion,
a deeper analysis. In what concerns the results of Ref. \cite{bean}, we
observe that the crossing of all observational data, including type Ia
supernovae, may put the Chaplygin gas again at a competitive level with
respect to $\Lambda CDM$. Our results indicate clearly that the Chaplygin
gas is preferred with respect to $\Lambda CDM$ if only type Ia supernovae
data is taken into account.

The $CGM$ predicts a Universe younger than $\Lambda CDM$, but still in
agreement with other astronomical data, in particular with the age of
globular clusters. Both models favour a closed Universe, with a very small
difference for the value of the curvature parameter. On the other hand, the
value of the Hubble parameter is essentially the same in both models, as
well as the deceleration parameter $q_{0}$, for which the data indicate a
highly negative value, near $-1$. The value of the deceleration parameter
becomes less negative for a flat Universe and when $CDM$ is absent.

In Ref. \cite{alcaniz}, statistics of gravitational lenses where used to
constrain the proportion of dark matter and Chaplygin gas in a flat
Universe, and the authors found $\Omega _{m0}\sim 0.2$. We remark also that
our results are consistent with the particular case described in Ref. \cite
{avelino}.

These results may be compared with those recently obtained from the $WMAP$
observatory for the spectrum of the anisotropy of the cosmic microwave
background radiation. Using also the $2dFGRS$ and Lyman-$\alpha $ forest
data, and fixing a $\Lambda CDM$ model, it has been obtained that $\Omega
_{m0}h^{2}=0.133\pm 0.006$, $h=0.72\pm 0.03$, where $H_{0}=100hMpc/km.s$,
and $\Omega _{k0}=0.02\pm 0.02$. We notice that the dark matter component is
much smaller than the value deduced from the supernovae data for $\Lambda
CDM $ ($\Omega _{m0}=0.58_{-0.58}^{+0.56}$), and at same time the Hubble
parameter is greater. The smaller uncertainty on the dark matter component
is natural in this case, since the CMB spectrum gives a better estimation on
the total matter in the Universe through the position of the first acoustic
peak, and it is verified that an almost flat Universe is favoured by the
data, for the $\Lambda CDM$ model, in contrast with what the supernovae data
indicate for the same theoretical framework.

For the $CGM$ there is a large uncertainty also in the estimation of the
curvature parameter if only supernovae data are used. But, its quite
remarkable that, if the dark matter density is fixed to zero in this model
the type Ia supernovae data indicates an almost flat Universe: $\Omega
_{k0}=0.17_{-1.58}^{+0.83}$. It must be stressed that a proper comparison
between the $WMAP$ results with our analysis for the Chaplygin gas requires
that the $WMAP$ data must be analysed using the $CGM$, what is one of the
natural extension of the present work.

We hope that, in the future, more SNe Ia data (from the SNAP project \cite
{linder1,linder2}, etc) with small observational errors will impose
stringent constraints on the parameter estimation for the $CGM$ and the $%
\Lambda CDM$ model, so the parameter credible regions become narrow enough
to rule out one of these cosmological models. For example, the estimation of
the parameter $\bar{A}$ could favour one of the models with high confidence
level ($>2\sigma $), the estimation of $\Omega _{m0}$ and $t_{0}$ could be
incompatible with other independent and well accepted estimations therefore
excluding some cosmological models, etc.

\vspace{0.5cm}

\vspace{0.5cm} \noindent \textbf{Acknowledgments} \newline
\newline
\noindent We would like to thank M. de Castro for his help on statistical
theory. During this work, the authors have received financial support from
CNPq, CAPES and FACITEC/PMV of Brazil.

\end{document}